\documentclass[twocolumn,showpacs,preprintnumbers,amsmath,amssymb,aps,prb,superscriptaddress]{revtex4}

\pdfoutput=1
\usepackage[pdftex]{graphicx}

\usepackage{dcolumn}% Align table columns on decimal point\right] 
\usepackage{bm}% bold math
\usepackage[caption=false]{subfig}% For figures side by side

\usepackage{color}
\usepackage{amstext}
\usepackage[utf8]{inputenc}% Text encoding

\newcommand{\bra}[1] {\left<#1\right|}
\newcommand{\ket}[1] {\left|#1\right>}
\newcommand{\braket}[2] {\left<#1|#2\right>}

\newcommand{\e}[1] {\text{e}^{#1}}
\newcommand{\I}[0] {\text{i}}

\newcommand*{\supp}[1]{\text{supp}\,#1}

\begin{document}

%%%%%%%%%%%%%%%%%%%%%%%%%%%%%%%%%%%%%%%%%%%%%%%%%%%%%%%%%%%%%%%%%%%%%%%%%%
%                               Title                                    %
%%%%%%%%%%%%%%%%%%%%%%%%%%%%%%%%%%%%%%%%%%%%%%%%%%%%%%%%%%%%%%%%%%%%%%%%%%
\title{Conductance through geometrically frustrated itinerant electronic systems}

\author{A. A. Lopes}
\affiliation{Institute of Physics, University of Freiburg, Hermann-Herder-Straße 3, 79104 Freiburg, Germany.}

\author{B. A. Z. António}
\affiliation{Department of Physics, I3N, University of Aveiro,\\
Campus de Santiago, Portugal}

\author{R. G. Dias}
\affiliation{Department of Physics, I3N, University of Aveiro,\\
Campus de Santiago, Portugal}

\date{\today}

%%%%%%%%%%%%%%%%%%%%%%%%%%%%%%%%%%%%%%%%%%%%%%%%%%%%%%%%%%%%%%%%%%%%%%%%%%
%                              abstract                                  %
%%%%%%%%%%%%%%%%%%%%%%%%%%%%%%%%%%%%%%%%%%%%%%%%%%%%%%%%%%%%%%%%%%%%%%%%%%
\begin{abstract}
We study a two terminal electronic conductance through an AB$_2$ ring which is an example of the family of itinerant geometrically frustrated electronic systems. These systems are characterized by the existence of localized states with nodes in the probability density. We show that such states lead to distinct features in the conductance. For zero magnetic flux, the localized states act as a filter of the zero frequency conductance peak,  if the contact sites have hopping probability to sites which are not nodes of the localized states. For finite flux, and in a chosen orthonormal basis, the localized states have extensions ranging from two unit cells to the complete ring, except for very particular values of magnetic flux. The conductance exhibits a zero frequency peak with a dip which is a distinct fingerprint of the variable extension of these localized states.

\end{abstract}

\maketitle

%%%%%%%%%%%%%%%%%%%%%%%%%%%%%%%%%%%%%%%%%%%%%%%%%%%%%%%%%%%%%%%%%%%%%%%%%%
%                             Introduction                               %
%%%%%%%%%%%%%%%%%%%%%%%%%%%%%%%%%%%%%%%%%%%%%%%%%%%%%%%%%%%%%%%%%%%%%%%%%%
\section{Introduction}
The conductance through molecular devices, nanowires and other nano systems has been extensively studied both theoretically and experimentally. 
Nano transport phenomena such as 
Coulomb blockade \cite{gorter_possible_1951}, 
conductance quantization \cite{van_wees_quantized_1988}, resonant tunneling \cite{stone_effect_1985}, quantum interference, Aharonov-Bohm oscillations in the conductance \cite{aharonov_significance_1959,webb_observation_1985} 
are now well understood.

The conductance fingerprints of localized states, however, induced by the topology of a nanocluster \cite{gulacsi_exact_2007,kikuchi_experimental_2005,macedo_magnetism_1995,montenegro-filho_doped_2006,vidal_aharonov-bohm_1998,tamura_flat-band_2002,tasaki_nagaokas_1998,mielke_ferromagnetism_1999,derzhko_structural_2005,richter_magnetic-field_2004,derzhko_low-temperature_2010,tanaka_metallic_2007,duan_theoretical_2001,gulacsi_exact_2005,gulacsi_exact_2003,richter_exact_2004,rule_nature_2008,schulenburg_macroscopic_2002,wu_flat_2007,wu_p_xy-orbital_2008,zhitomirsky_lattice_2007} 
has never been addressed as far as we know.
Do these localized states inhibit the electronic transport through the cluster or is the conductance indifferent to their existence?
The answer is rather complex and unexpected.
In this paper we show that, in the case of the AB$_2$ ring (which is an example of the family of itinerant geometrically frustrated electronic systems \cite{gulacsi_exact_2007,kikuchi_experimental_2005,macedo_magnetism_1995,montenegro-filho_doped_2006,vidal_aharonov-bohm_1998,tamura_flat-band_2002,tasaki_nagaokas_1998,mielke_ferromagnetism_1999,derzhko_structural_2005,richter_magnetic-field_2004,derzhko_low-temperature_2010,tanaka_metallic_2007,duan_theoretical_2001,gulacsi_exact_2005,gulacsi_exact_2003,richter_exact_2004,rule_nature_2008,schulenburg_macroscopic_2002,wu_flat_2007,wu_p_xy-orbital_2008,zhitomirsky_lattice_2007}),
these localized states  act as zero frequency conductance absorbers for zero magnetic flux, but  surprisingly generate a dipped zero frequency conductance peak when magnetic flux is applied.
Similar features should be observed in the conductance  through other elements of the family of the itinerant geometrically frustrated electronic systems of the Lieb lattice kind, that is, systems which display localized states with nodes in their probability  density\cite{lopes_interacting_2011}.

This paper is organized in the following way: First we recall recent exact results about the eigenstates of the AB$_2$ tight-binding ring, and in particular we discuss the form of the localized states when magnetic flux is present.  Next we discuss the conductance through an AB$_2$ ring for several different scenarios including different placements of the conducting leads and different values of the threading flux.

%%%%%%%%%%%%%%%%%%%%%%%%%%%%%%%%%%%%%%%%%%%%%%%%%%%%%%%%%%%%%%%%%%%%%%
%                       AB$_2$ chain                                 %
%%%%%%%%%%%%%%%%%%%%%%%%%%%%%%%%%%%%%%%%%%%%%%%%%%%%%%%%%%%%%%%%%%%%%%
\section{Exact diagonalization of the AB$_2$ chain and localized states}
\begin{figure}[th]
    \centering
    \includegraphics[width=6cm]{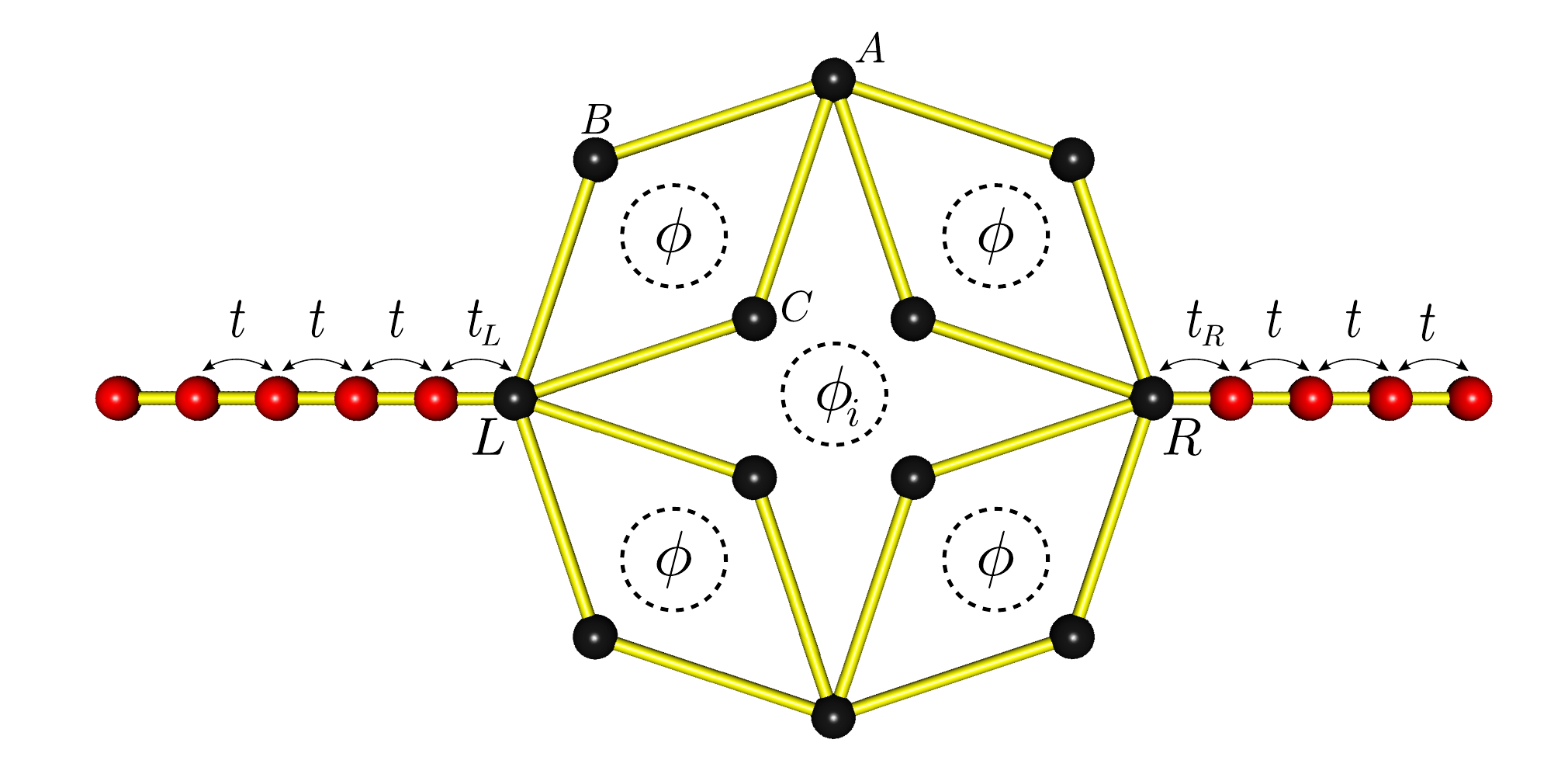}
    \caption{The AB$_2$ ring is connected at sites $L$ and $R$, to semi-infinite tight binding leads via a hopping amplitude $t'$. Except where otherwise stated, the hopping amplitude of the leads is taken to be the same to that of the star, $t$. Here it is shown the situation for $N_c = 4$, a particular case we will study in detail. The magnetic fluxes threading the plaquettes and the inner ring are respectively $\phi$ and $\phi_i$.}
    \label{fig:DiamondStar4SidedLeads}
\end{figure}

In order to address the phenomena of coherent transport through an AB$_2$ ring, we consider a two terminal set up of one-dimensional (1D) tight-binding leads coupled to the AB$_2$ ring, as depicted in Fig.~\ref{fig:DiamondStar4SidedLeads}. Our results are easily generalized  to the case of 3D leads as long as only one site of each lead  contacts the cluster. 
We shall often focus on the case where the number of cells of the  AB$_2$ ring, $N_c$, is equal to 4, and assume that each plaquette is threaded by an identical magnetic flux, $\phi$.

%%%%%%%%%%%%%%%%%%%%% 
%      figure       % 
%%%%%%%%%%%%%%%%%%%%%
%%%%%%%%%%%%%%%%%%%%% 
%      end          % 
%%%%%%%%%%%%%%%%%%%%%

%%%%%%%%%%%%%%%%%%%%
%      figure       %
%%%%%%%%%%%%%%%%%%%%%
\begin{figure*}[ht]
\centering
\begin{minipage}{7cm}
    \subfloat[]{\label{fig:decoupling}\includegraphics[width=1\textwidth]{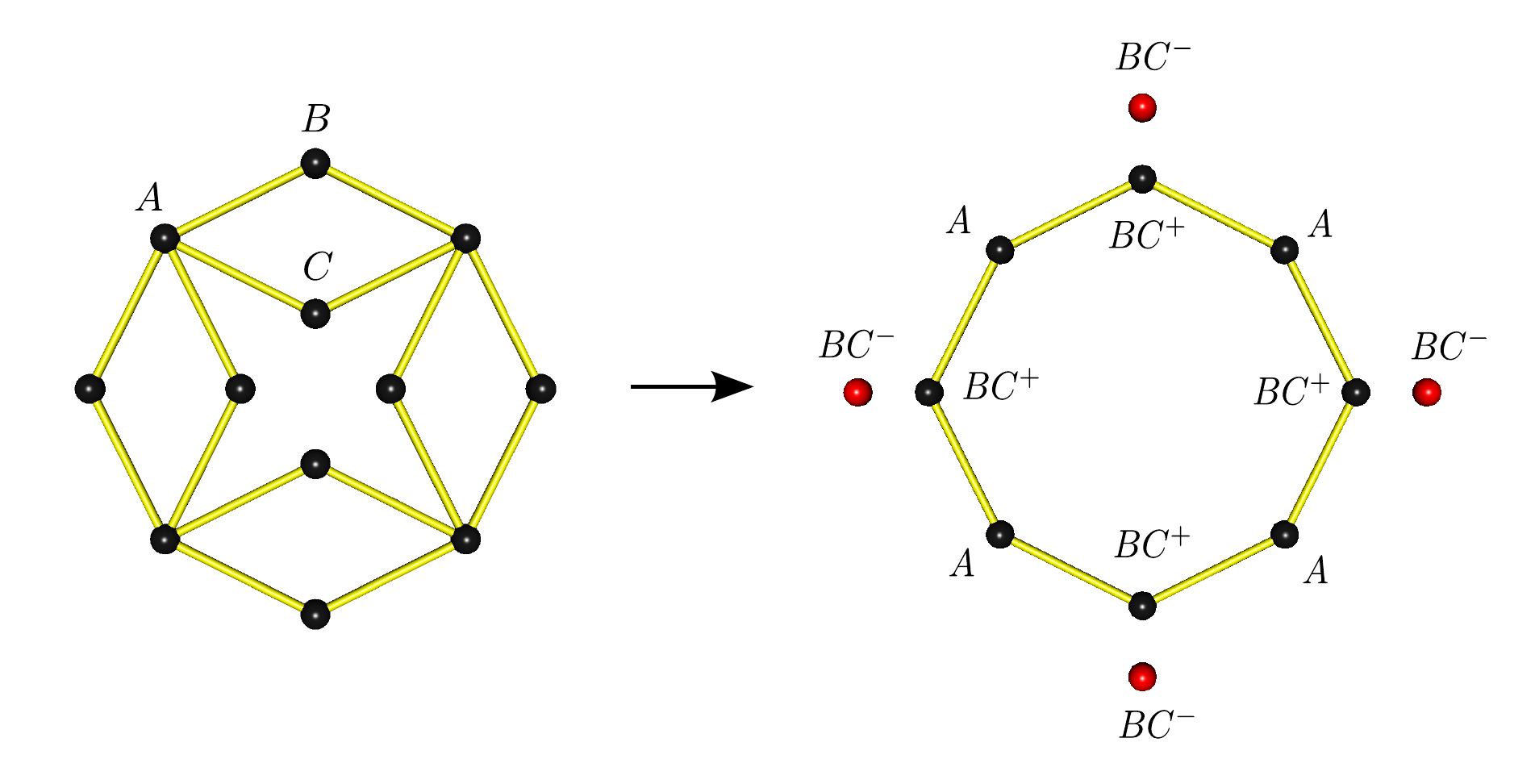}}  
\end{minipage}
\begin{minipage}{7cm}
  \subfloat[]{\label{fig:figBC}\includegraphics[width=0.8\textwidth]{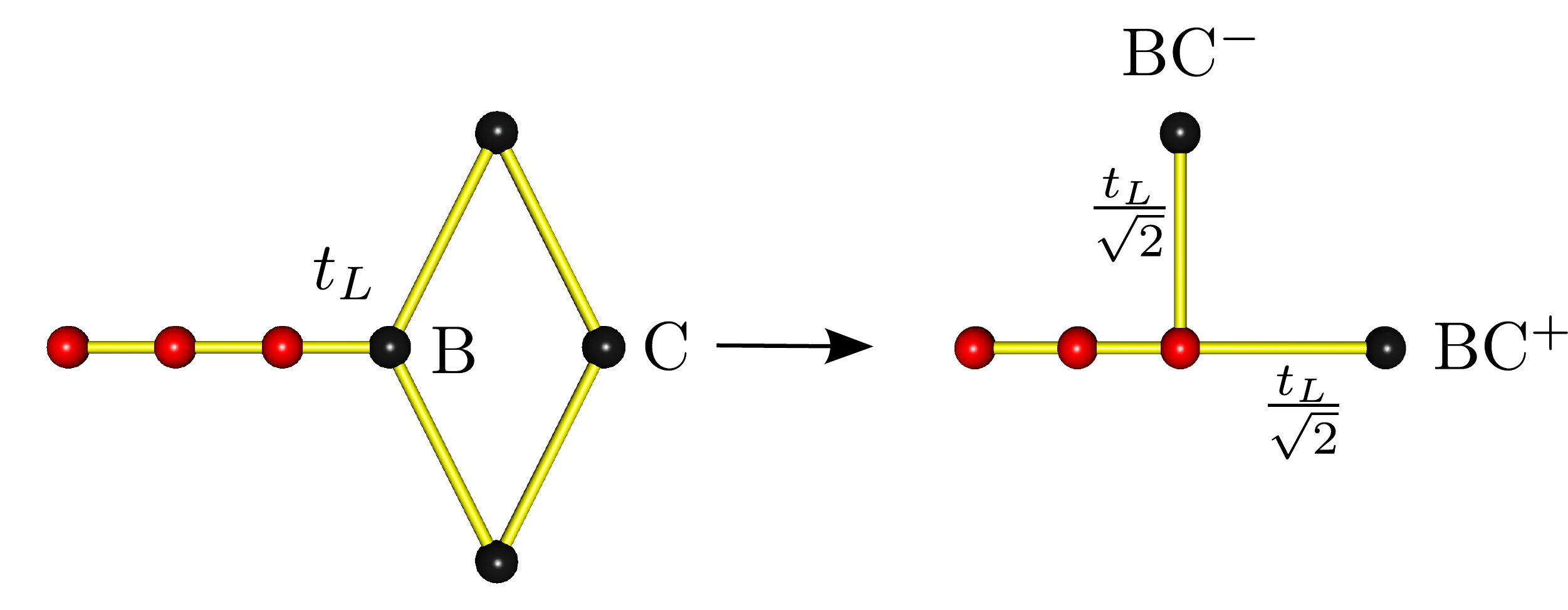}}  \\
  \hspace{.2cm}\subfloat[]{\label{fig:figonelink}\includegraphics[width=0.8\textwidth]{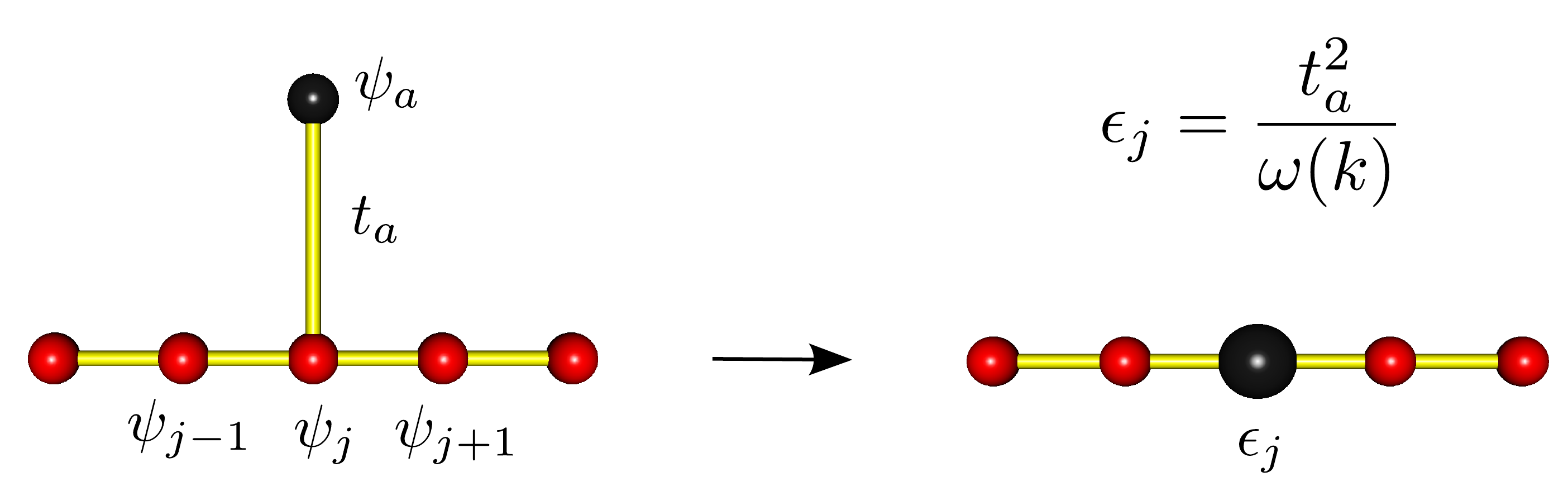}} 
\end{minipage}
\caption{(a) In the absence of magnetic flux, rewriting  the Hamiltonian in the basis of antibonding BC$^-$, bonding BC$^+$ and  A states, one obtains a tight-binding ring of sites A and bonding BC sites (with hopping constant $\sqrt{2} t$) and a ring of decoupled anti-bonding BC states. (b) The hopping term from the left lead to a B site of the AB$_2$ ring, in the basis of antibonding BC, bonding BC and  A states, becomes $t_L/\sqrt{2}$. (c) For a incident particle with energy $\omega=-2t \cos (k)$, an extra transverse hopping $t_a$  to a dangling site effectively modifies the on-site energy of site $j$ to $\epsilon_j = t_a^2/\omega$. }  
\label{fig:decoupling2}
\end{figure*}
%%%%%%%%%%%%%%%%%%%%%
%      end          %
%%%%%%%%%%%%%%%%%%%%%

\subsection{Exact diagonalization}
The Hamiltonian of the full system is given by
\begin{equation}
    H_\text{ring} + H_\text{leads} + H_\text{LR},
\end{equation}
where $H_\text{leads}$ is the Hamiltonian of the isolated leads, assumed to be semi-infinite,
%\begin{equation}
%    H_\text{leads} = -t \sum_{j=1,\sigma}^\infty a_{j,\sigma}^\dagger a_{j+1,\sigma} + a_{-j-1,\sigma}^\dagger a_{-j,\sigma} + \text{H.c.},
%\end{equation}
\begin{equation}
    \begin{split}
    H_\text{leads} &= -t \sum_{j=1}^\infty \sum_{\sigma = \uparrow, \downarrow} \ket{a_{j,\sigma}}\bra{a_{j+1,\sigma}}  \\
    &+ \ket{a_{-j-1,\sigma}} \bra{a_{-j,\sigma}} + \text{H.c.},
    \end{split}
\end{equation}
where $\ket{a_{j},\sigma}$ is a lead Wannier state at site $j$ and with spin $\sigma$. $j \in (-\infty,-1]$ correspond to left lead states while $j \in [1,\infty)$ correspond to right lead states.
$H_\text{star}$ is the Hamiltonian for an AB$_2$ chain with $N_c$ unit cells,
%\begin{eqnarray}
%    H_\text{star} = -t \sum_{j=1}^{N_c} &  & \left[ \e{\I \phi_o/2N_c} ( A_j^\dagger B_j  + B_j^\dagger A_{j+1}) \right. \\
%    & + & \left. \e{-\I \phi_i/2N_c} ( C_j^\dagger A_j + A_{j+1}^\dagger C_j ) \right] + \text{H.c.}. \nonumber
%\end{eqnarray}
\begin{equation}
    \begin{split}
    H_\text{ring} &= -t \sum_{j=1}^{N_c} \sum_{\sigma = \uparrow, \downarrow} \e{\I \phi_o/2N_c} ( \ket{A_{j,\sigma}}\bra{B_{j,\sigma}}  + \ket{B_{j,\sigma}} \bra{A_{j+1,\sigma}}) \\
    & + \e{-\I \phi_i/2N_c} ( \ket{C_{j,\sigma}}\bra{A_{j,\sigma}} + \ket{A_{j+1,\sigma}}\bra{C_{j,\sigma}} ) + \text{H.c.}, \nonumber
    \end{split}
\end{equation}
where $\ket{A_{j,\sigma}}$, $\ket{B_{j,\sigma}}$, $\ket{C_{j,\sigma}}$ correspond to states on A, B and C sites, respectively, of the $j$th cell/plaquette, with spin $\sigma$.
Here we have chosen a gauge such that the Peierls phases are equally distributed in the inner ring and in the outer ring. 
In Fig.~\ref{fig:DiamondStar4SidedLeads}, an AB$_2$ ring is shown with a magnetic flux $\phi$ threading each plaquette and a magnetic flux $\phi_i$ threading the inner ring. The magnetic flux enclosed by the outer ring is 
$\phi_o = \phi_i + 4N_c \phi/4$ and we  introduce an auxiliary flux $\phi^\prime$ such that $\phi_o = \phi^\prime + 2N_c \phi/4$, $\phi_i = \phi^\prime - 2N_c \phi/4$.
 The  inner sites in the AB$_2$ ring of Fig.~\ref{fig:DiamondStar4SidedLeads} are denoted as C sites and the  outer sites as B sites. Spinal sites are denoted as A sites.
The hybridization between the AB$_2$ ring and the leads is given by 
\begin{equation}
    H_{\text{LR}} = - \sum_{\sigma = \uparrow, \downarrow} t_L \ket{a_{-1,\sigma}} \bra{X_{L,\sigma}} + t_R \ket{a_{1,\sigma}} \bra{X_{R,\sigma}} + \text{H.c.},
\end{equation}
where $t_{L,R}$ are the hopping amplitudes coupling the leads and the star and $X$ stands for an $A$, $B$ or $C$ site depending on where the left (L) and right (R) contacts are.
Since we don't consider spin-spin interactions, each spin channel is independent and we disregard spin in the rest of the paper, without any loss of generality.

%%%%%%%%%%%%%%%%%%%%% 
%      figure       % 
%%%%%%%%%%%%%%%%%%%%% 
\begin{figure}[tp]
    \includegraphics[width=.4 \textwidth]{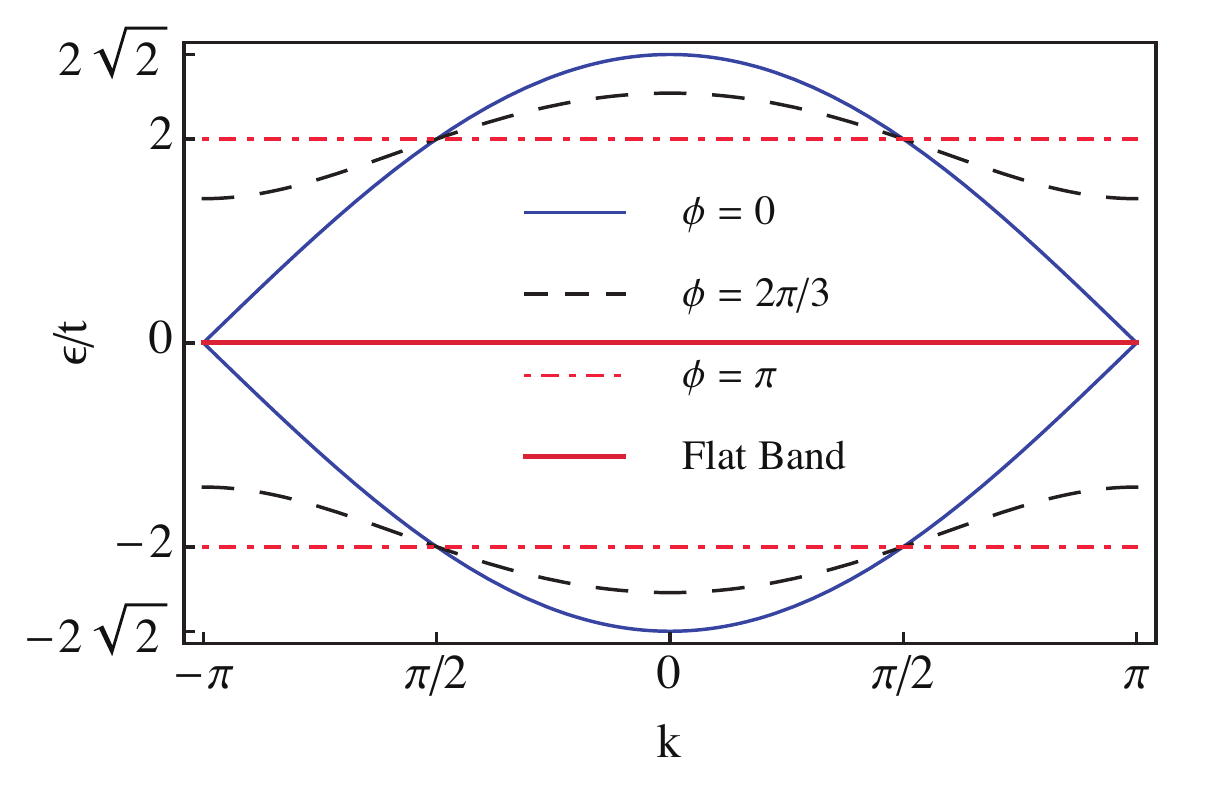} 
    \caption{Dispersion relation of the AB$_2$ ring for $\phi^\prime = 0$ and several values of $\phi$. Note that a gap opens between the localized band and the itinerant bands when there is a finite flux. Also, for $\phi = \pi$ all bands are flat, and therefore, all states are localized.}
    \label{fig:dispersion}
\end{figure}
%%%%%%%%%%%%%%%%%%%%% 
%      end          % 
%%%%%%%%%%%%%%%%%%%%%
Without interactions, the tight-binding AB$_2$ chain has a flat band even in the presence of magnetic flux (Fig.~\ref{fig:dispersion} displays the dispersion relation of the nearest neighbor AB$_2$ chain for several values of the plaquette threading flux).
The eigenvalues for an arbitrary value of flux are  given by
\begin{equation}
        \begin{split}
            \epsilon_\text{flat} &= 0, \\
            \epsilon_{\pm} &= \pm 2t \sqrt{1 + \cos(\phi/2)\cos(\phi'/N_c + k)},
        \end{split}
\end{equation}
where $k$ is the momentum.

\subsection{Localized states}
Localized states associated with the flat band can be written in the most compact form as standing waves in one (in the absence of magnetic flux) or two consecutive plaquettes (in the presence of magnetic flux).\cite{lopes_interacting_2011} 
In the particular case of  zero flux,  localized states are simply the anti-bonding combination of the B and C states, 
BC$_j^-= ( \ket{B_{j}}- \ket{C_{j}})/\sqrt{2 }$, and  itinerant states in the AB$_2$ ring are linear combinations of A and bonding BC$^+$ states, BC$_j^+= ( \ket{B_{j}} + \ket{C_{j}})/\sqrt{2 }$.
Rewriting  the Hamiltonian in the basis of antibonding BC$^-$, bonding BC$^+$ and  A states, one obtains a tight-binding ring of sites A and bonding BC sites (with hopping constant $\sqrt{2} t$) and a ring of decoupled anti-bonding BC states,\cite{lopes_interacting_2011} as shown in  Fig.~\ref{fig:decoupling}.

%%%%%%%%%%%%%%%%%%%%% 
%      figure       % 
%%%%%%%%%%%%%%%%%%%%% 
\begin{figure}[t]
\centering
	\subfloat[$\phi = 0$]{\includegraphics[height=2.cm]{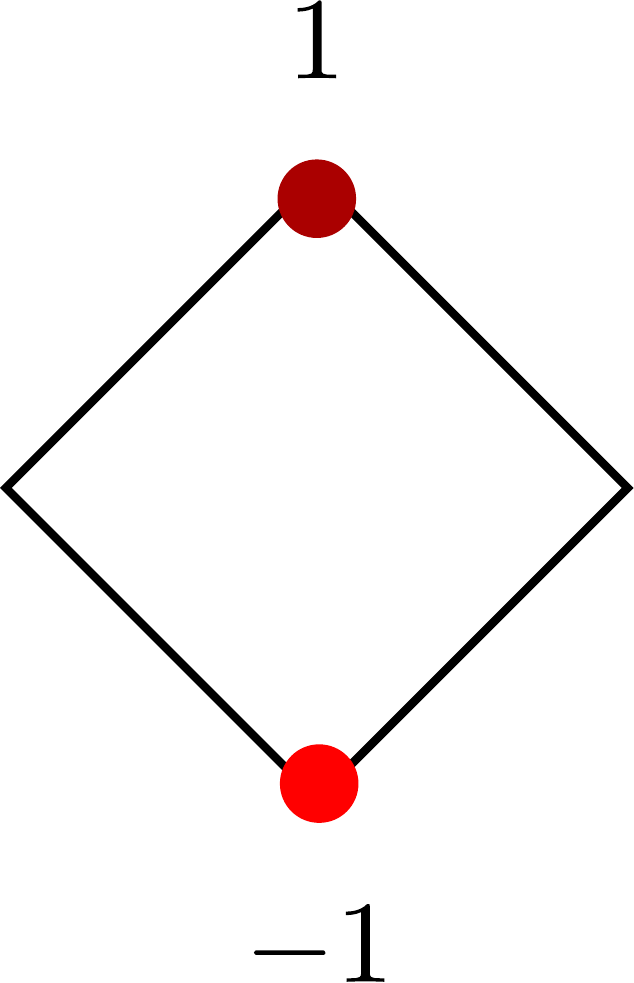}
	\label{fig:loczeroFlux}}
	\quad
	\subfloat[Ferromagnetic]{\includegraphics[height=2.cm]{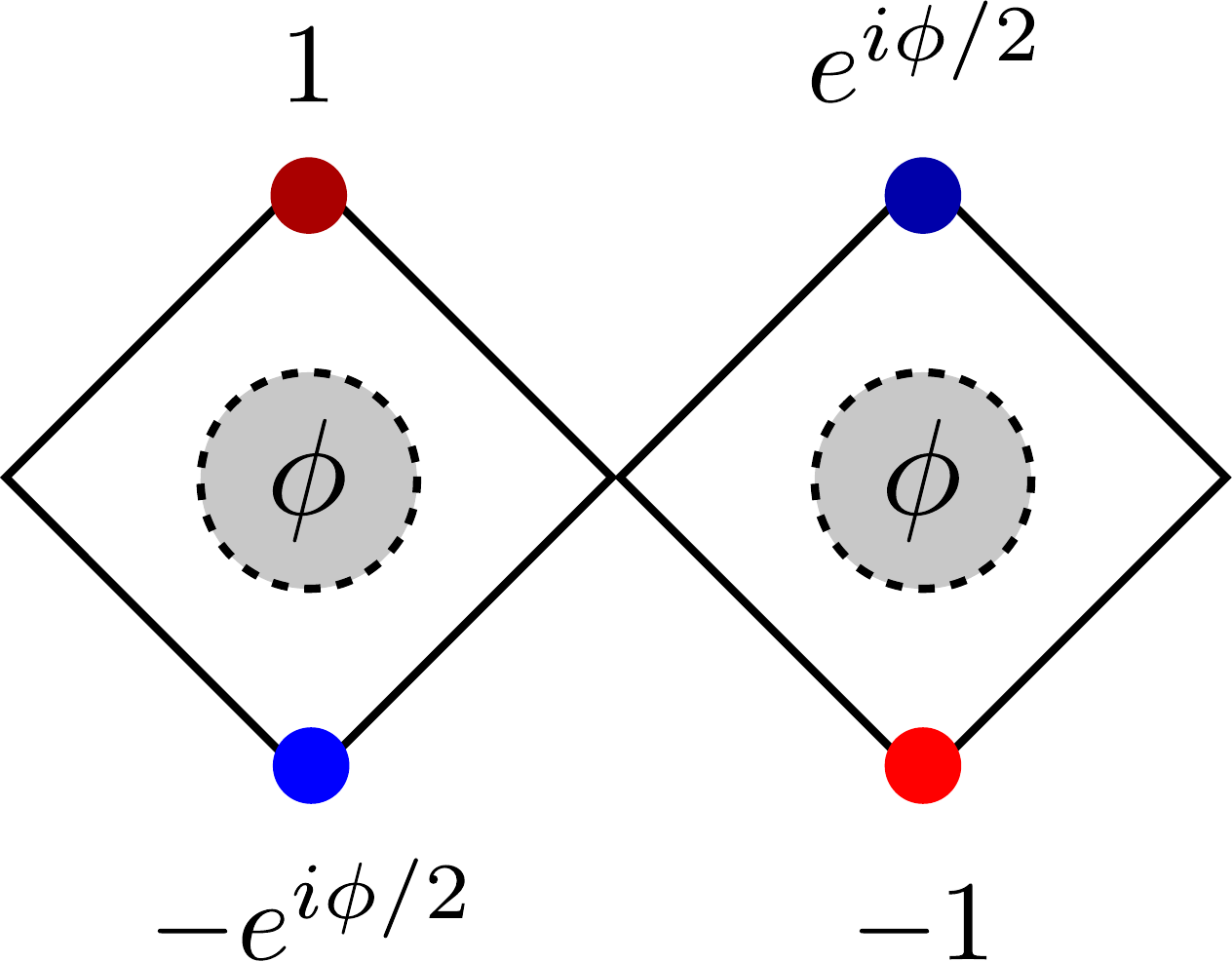}
	\label{fig:locSymm}}
	\quad
	\subfloat[Antiferromagnetic]{\includegraphics[height=2.cm]{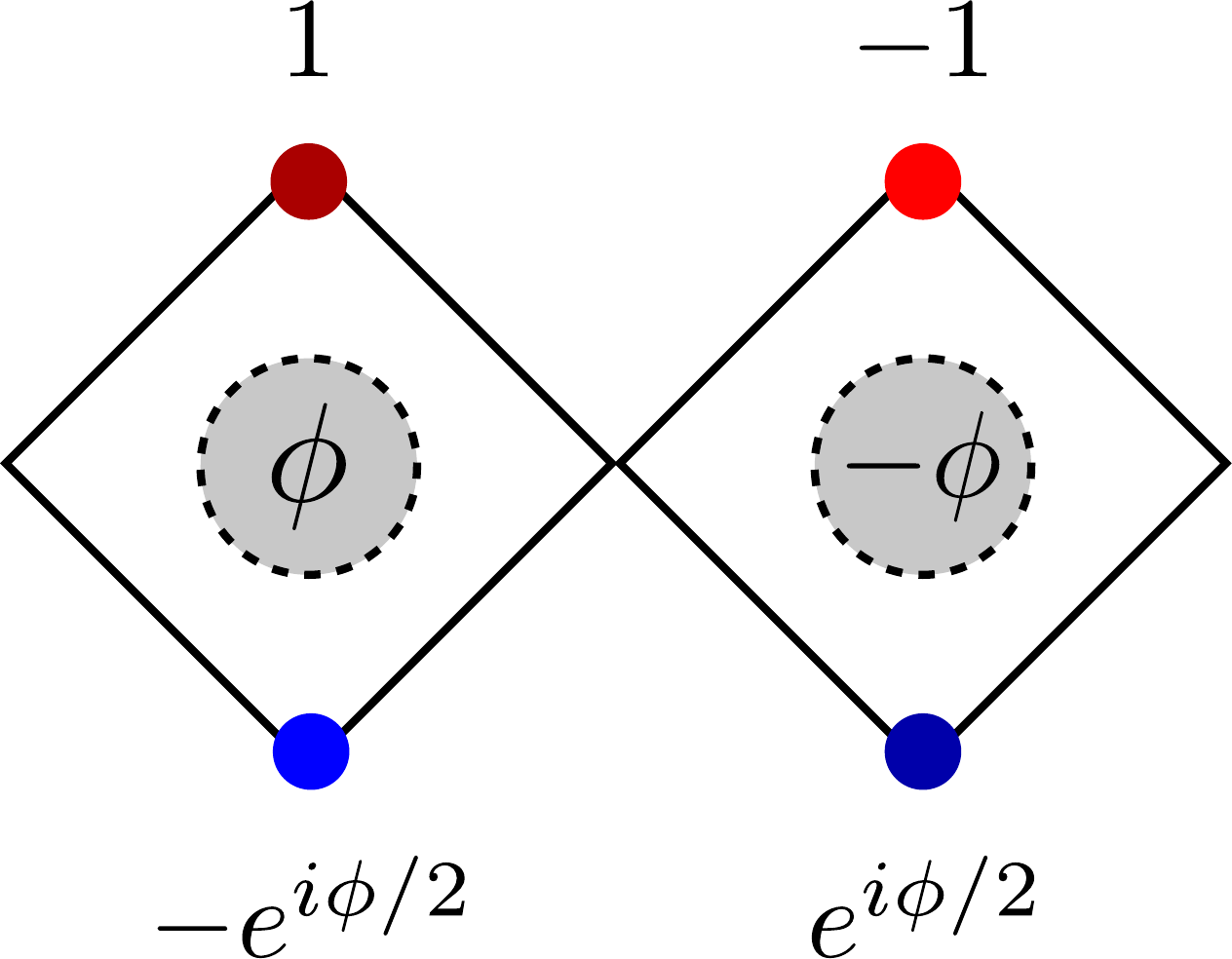}
	\label{fig:locASymm}} \\
	\subfloat[Two cells periodic flux]{\includegraphics[height=2.2cm]{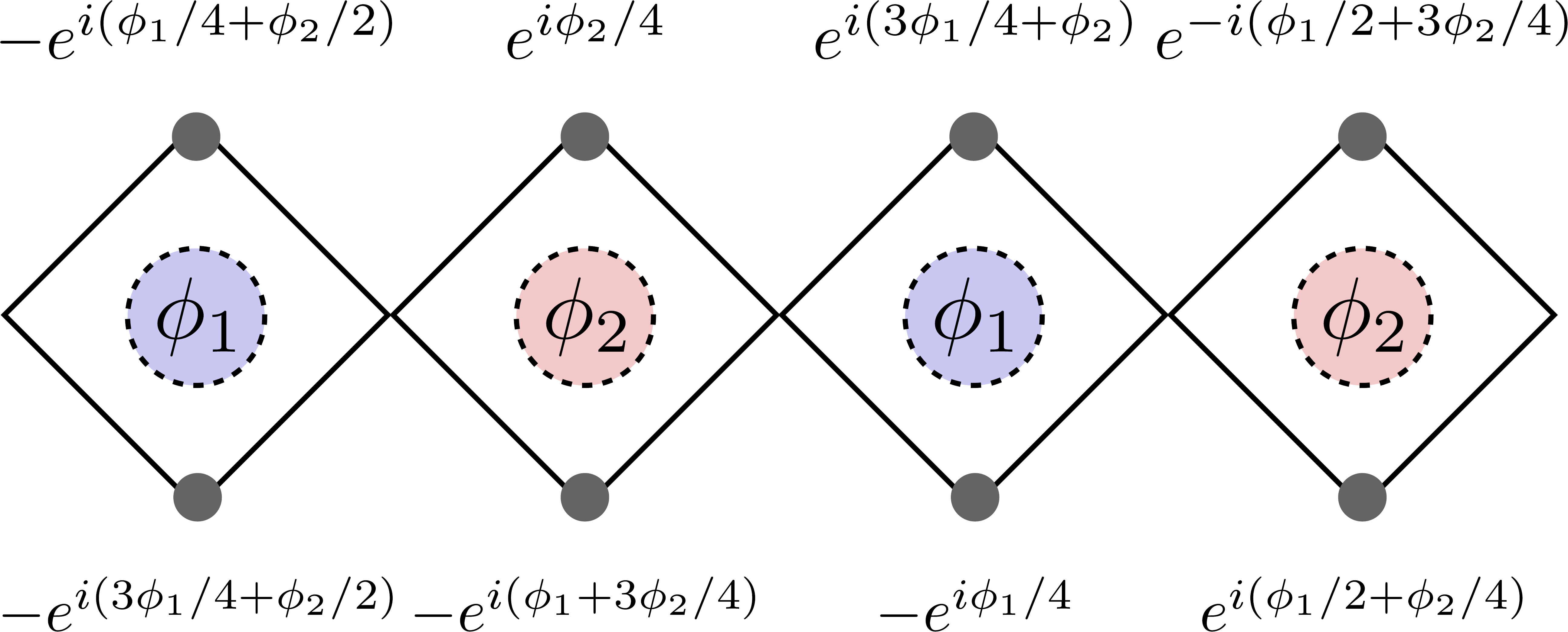}
	\label{fig:twoFluxes}} \\
    \subfloat[Localized states orthonormalization]{\includegraphics[height=4cm]{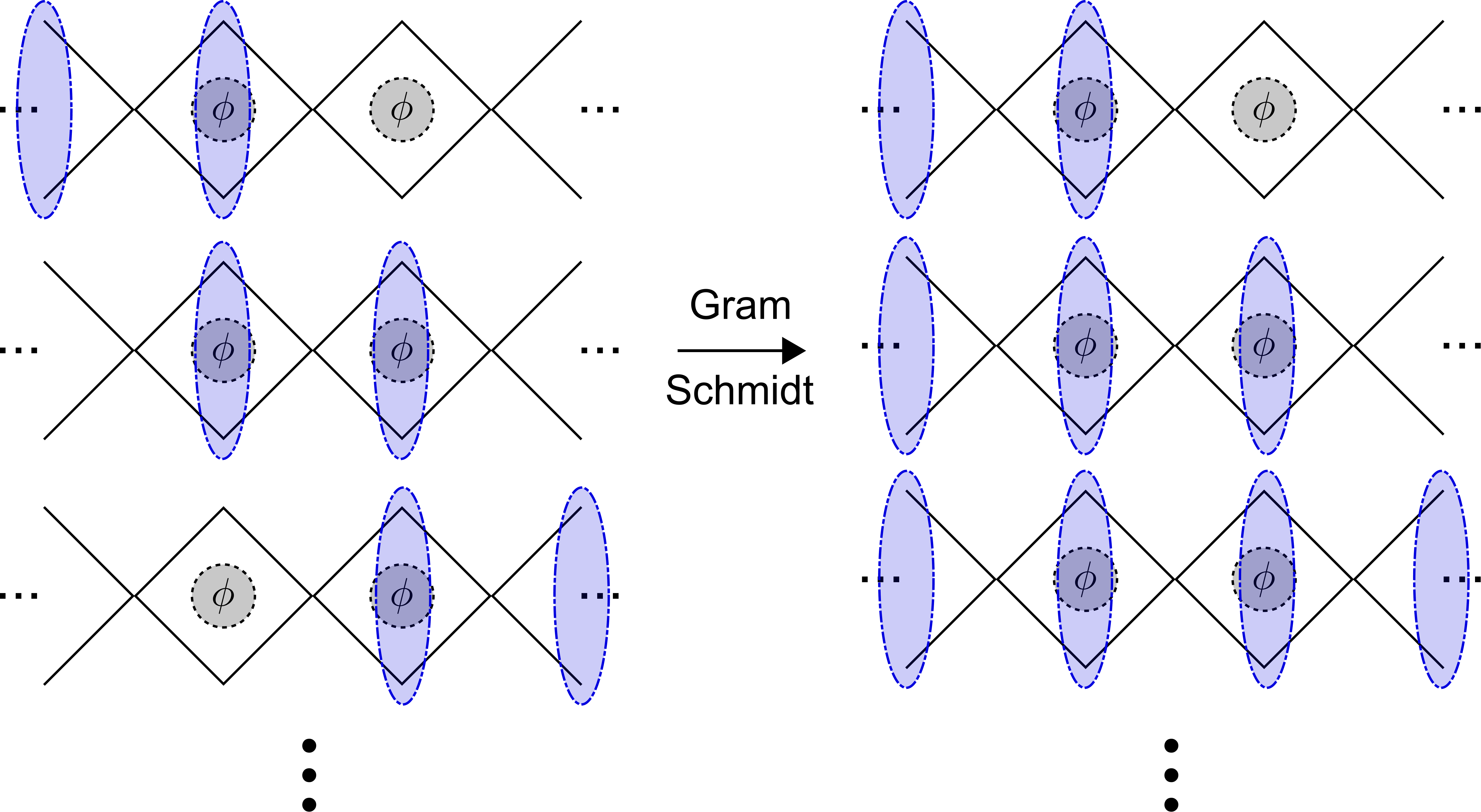}
	\label{fig:gramSchmidt}}
    \caption{(a), (b) and (c): Localized states for AB$_2$ chains without flux and threaded by ferromagnetic or anti-ferromagnetic flux, respectively. For simplicity the states are not normalized and we only draw the cells where the wavefunction is non-zero. (d) Localized states for $N=2$, where $N$ is the periodicity in the flux (in terms of number of cells). (e) Our non-orthonormal basis of localized states, occupying two cells, can be transformed into an orthonormal basis where they occupy, $1, 2, \ldots N_c$ cells, via the Gram-Schmidt procedure.}
    \label{fig:locStates}
\end{figure}
%%%%%%%%%%%%%%%%%%%%% 
%      end          % 
%%%%%%%%%%%%%%%%%%%%%

The number of localized states is equal to the number of rhombi and in the presence of flux, if written in the most compact form (each localized state taking place in two consecutive plaquettes) they form a non-orthogonal set of states. Orthogonalization of these set of states implies that the extension of the localized states ranges from two consecutive plaquettes to  the complete ring (see Fig.~\ref{fig:gramSchmidt}), except for $\phi=0$ 
%(in this case, the orthogonal localized states written in the most compact form occupy one plaquette ) 
and for $\phi=\pi$  (in this case, the orthogonal localized states occupy only two   consecutive plaquettes).
This will imply a clear difference in the conductance when compared with the zero flux case. Note that a gap opens between the localized band and the itinerant bands when flux is present.

Assuming $\phi'=0$ to simplify, the non-orthogonal localized states are of the form 
$
	(\ket{B_j}
	- e^{i\frac{\phi}{2}} \ket{C_j})
	+ (e^{i\frac{\phi}{2}} \ket{B_{j+1}}
		- \ket{C_{j+1}})
$
where the sites have been numbered clockwise in the AB$_2$ ring, that is, $j$ indexes the plaquettes in the AB$_2$ ring.

Since the localized states have nodes at the A sites, we can write these localized states indicating only the one-particle state amplitudes at the pairs of B and C sites of the AB$_2$ ring, 
that is, we can write the localized state as a list with $2N_c$ entries $(b_1,c_1 \ldots, b_n, c_n)$, where $b_j$ and $c_j$ are, respectively, the value of the wavefunction on site $B_j$ and $C_j$.
Then our localized states are,
\begin{equation}
\ket{\psi_j} = \dfrac{1}{\sqrt{4}} \left(
 0, \ldots, 0,\underbrace{1}_{b_j}, \underbrace{-\e{-\I \phi/2}}_{c_j}, \underbrace{\e{-\I \phi/2}}_{b_{j+1}}, \underbrace{-1}_{c_{j+1}}, 0, \ldots, 0 \right)
\end{equation}
Note that for $\phi=0$, we have for $\ket{\psi_j}$ that $b_j = b_{j+1}$ and $c_j = c_{j+1}$, and 
and for  $\phi=\pi$, $\braket{\psi_{j}}{\psi_{j+1}}=0$.
There are many possible ways of constructing an orthogonal basis for the subspace of localized states. Our results for the conductance are obviously independent of this choice.
We simply use the Gram-Schmidt orthogonalization, starting with the basis
\begin{equation}
    \begin{split}
        \ket{\psi_1} &= \dfrac{1}{\sqrt{4}}\left( 
        1, -\e{-\I \phi/2}, \e{-\I \phi/2}, -1,
        0, \ldots, 0 \right), \\
        \ket{\psi_2} &= \dfrac{1}{\sqrt{4}}\left( 
        0, 1, -\e{-\I \phi/2}, \e{-\I \phi/2}, -1, 
        0, \ldots, 0 \right), \\
                     &\vdots \\
        \ket{\psi_{N_c}} &= \dfrac{1}{\sqrt{4}}\left( 
               \e{-\I \phi/2}, -1,
                0, \ldots, 0, 1, -\e{-\I \phi/2} \right).                   
    \end{split}
\end{equation}
For such a basis we have
\begin{equation}
    \braket{\psi_i}{\psi_j} = \delta_{i,j} + \dfrac{\cos \phi/2 }{2} \left( \delta_{i-1,j} + \delta_{i+1,j}\right).
\end{equation}
For simplicity, let us define the support of a wavefunction, denoted by $\supp$, to be those site where the wavefunction is non-zero. Then we have $\supp {\ket{\psi_j}} = \lbrace B_j, C_j, B_{j+1}, C_{j+1} \rbrace$.

Let $\lbrace \ket{\phi_j} \rbrace_{j=1}^{N_c}$ denote the orthonormalized basis after the G-S procedure,
defined by the recursive expression,

\begin{equation}
    \begin{split}
        \ket{\phi^\prime_j} &= \ket{\psi_j} - \sum_{i=1}^{j-1}
        \braket{\phi_i}{\psi_j}\ket{\phi_i}, \\
        \ket{\phi_j} &= 
            \dfrac
            {
              \ket{\phi_j^\prime}
            }{
              \sqrt
              {
                \braket
                {
                  \phi_j^\prime
                }{
                  \phi_j^\prime
                 }
                }
              }.
    \end{split}
\end{equation}

We focus on $\phi \neq \pi$ (for in that case, the basis is already orthonormalized) and begin by making $\ket{\phi_1} = \ket{\psi_1}$, which implies $\supp{\phi_1} = \lbrace B_1, C_1, B_2, C_2 \rbrace$. Then $\ket{\phi_2^\prime} = \ket{\psi_2} - \braket{\phi_1}{\psi_2}\ket{\phi_1}$. In this case, since $\braket{\phi_1}{\psi_2} \neq 0$, 
$\supp \phi_2 = \lbrace B_1, C_1, B_2, C_2, B_3, C_3 \rbrace$.
We then have $\ket{\phi_3^\prime} = \ket{\psi_3} - \braket{\phi_2}{\psi_3}\ket{\phi_2} - \braket{\phi_1}{\psi_3}\ket{\phi_1}$. Note that $\ket{\phi_1}$ and $\ket{\psi_3}$ have disjoint support, hence $\braket{\phi_1}{\psi_3} = 0$. Also, $\braket{\phi_2}{\psi_3} \propto \braket{\psi_2}{\psi_3} \neq 0$. Since $\supp{\ket{\psi_3}} = \lbrace B_3, C_3, B_4, C_4 \rbrace$ and $\supp{\phi_2} = \lbrace B_1, C_1, \ldots B_3, C_3 \rbrace$, and since, there is no destructive interference on sites $B_3$ and $C_3$ (it is a simple exercise to show this), $\supp{\phi_3} = \lbrace B_1, C_1, \ldots B_4, C_4 \rbrace$.
Continuing the above procedure we finally arrive at
\begin{equation}
    \supp{\ket{\phi_j}} = \lbrace B_1, C_1, \ldots B_{j+1}, C_{j+1} \rbrace.
\end{equation} 
and therefore the extension of the orthogonalized localized states (constructed this way) ranges between two consecutive plaquettes and the full AB$_2$ ring. This is illustrated schematically in  Fig.~\ref{fig:gramSchmidt}.

However, as we have already mentioned, one has two exceptions: 

i) for   $\phi=0$, 
the states $\ket{\alpha_j} = 
(0, \cdots,0, \underbrace{1}_{b_j}, \underbrace{-1}_{c_j} , 0, \cdots,0)$ already constitute an orthogonal set of localized states for $\phi=0$ as stated in the previous paragraph; 

ii) for  $\phi=\pi$, $\braket{\psi_{j}}{\psi_{j+1}}=0$ are orthogonal and in this case the range of the localized states in their most compact form is just two plaquettes.

There are many possible ways of constructing an orthogonal basis for the subspace of localized states. It must be stressed that our results for the conductance are obviously independent of this choice.

Using the construction for localized states of \cite{lopes_interacting_2011}, it is easy to extend some of the results presented in this paper to geometries other than the AB$_2$ geometry.
To make this more concrete let us give some examples. Let us start by considering an AB$_2$ chain with an arbitrary number of cells. Assume, for now, that the flux through each cell has the same value, a situation we call \emph{ferromagnetic} (shown in  Fig.~\ref{fig:locSymm}). Then, localized states occupying only two cells can be found for an arbitrary value of flux (albeit non orthogonal, except when  $\phi = \pi$),\cite{lopes_interacting_2011}  while for zero flux one can find localized states occupying only one cell as shown in  Fig.~\ref{fig:loczeroFlux}.
Now consider a situation where the magnetic flux through each plaquette is symmetric to the one threading its neighboring cells, a situation we call \emph{anti-ferromagnetic}. Then a similar state to the situation above can be found as is shown in  Fig.~\ref{fig:locASymm}.
For this particular case, using this construction, we can find localized states that occupy two cells. However, these states form an orthonormal basis only for $\phi = \pi$ and for the ferromagnetic flux situation, as can be readily seen calculating the overlap between neighboring states. Let $\ket{\psi_j}$ be the  state localized in the $j$th and $(j+1)$th cells. For the ferromagnetic situation the overlap between neighboring states is $\braket{\psi_{j+1}}{\psi_j} = \dfrac{\cos(\phi/2)}{2}$ while for the anti-ferromagnetic situation one has
$\braket{\psi_{j+1}}{\psi_j} = - \dfrac{1}{2}$.
Note that this flux threading each cell is not necessarily an external flux, since it may be generated by the spin of an atom/molecule, embedded into the chain as is the case of some coper oxide systems, namely CuO$_4$ chains \cite{matsuda_ordering_1998,schlappa_spin-orbital_2012}.

If the flux through each plaquette is distinct, but repeats every $N$ cells (Fig.~\ref{fig:twoFluxes} shows the situation for $N=2$), one can also use the same construction to find localized states. In this case however, one must consider $2N$ adjacent cells instead of 2, as before, and we will find a localized states that extend through $2N$ cells. As before, these are not necessarily orthonormal, but can be made so by using Gram-Schmidt orthonormalization. In the  extreme case, where there is no periodicity, translational invariance is obviously broken and our procedure will give us an extended state instead.
A particular case of the $N=2$ situation, with $\phi_1 = 2 \phi_2$ has been studied in \cite{movilla_quantum_2011}.

%%%%%%%%%%%%%%%%%%%%%%%%%%%%%%%%%%%%%%%%%%%%%%%%%%%%%%%%%%%%%%%%%%%%%%%%%%
%                       Conductance in the diamond star                  %
%%%%%%%%%%%%%%%%%%%%%%%%%%%%%%%%%%%%%%%%%%%%%%%%%%%%%%%%%%%%%%%%%%%%%%%%%%
\section{Conductance through the AB$_2$ ring} 
 
In this section we discuss the conductance through the AB$_2$ ring. We will begin by addressing the case without magnetic flux. Since no two-particle interactions are considered in this paper, the transmission probability $|t(\omega)|^2 $ for an incident particle with momentum $k$ and energy $\omega=-2 \cos (k)$  can be calculated using quantum scattering theory \cite{taylor_scattering_2006}, and 
is given by the 
following expression \cite{enss_impurity_2005},
\begin{eqnarray}
    \label{condu4}
    |t(\omega)|^2
    &=& 4t^2_Lt^2_R \sin ^2k 
    |\langle R\vert [\epsilon _k {\hat I_s}-H_s \nonumber \\
    & +& 
     e^{ik}
    \left(t_L^2\vert L\rangle\langle L\vert +t_R^2\vert R\rangle 
    \langle R \vert \right)]^{-1} \vert L\rangle|^2  ,
\end{eqnarray}
where the inverse is to be found within the subspace of the cluster sites (in our case, the AB$_2$ ring) 
positions and ${\hat I_s}$ is the identity operator in that subspace. 
This equation includes  the effect of the coupling of the ring to the leads as modifications of the on-site energies of sites $L$ and $R$.
If the conductance is normalized by the conductance of an ideal one dimensional system, $G_0 = e^2/\pi \hbar$, then the conductance is given by the transmission probability at the chemical potential. \cite{imry_conductance_1999}. In what follows, we will always deal with this normalized conductance, i.e., transmission probability.

%%%%%%%%%%%%%%%%%%%%% 
%      figure       % 
%%%%%%%%%%%%%%%%%%%%% 
\begin{figure*}[ht]
    \centering
    \includegraphics[width=1 \textwidth]{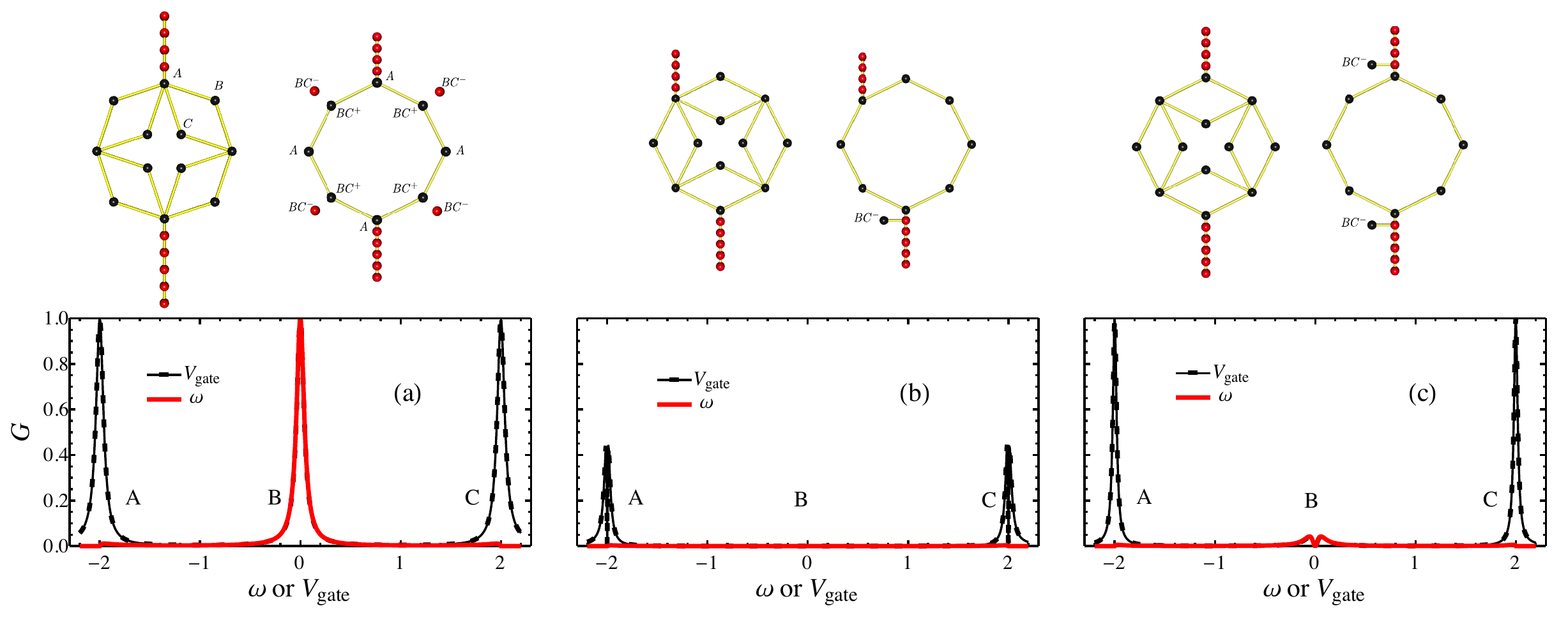}
    \caption{Normalized conductance through the AB$_2$ ring as function of the energy of the incident electron (or chemical potential of the leads) and as function of $V_\text{gate}$ for several positions of the leads. The positions of the leads are shown on the top figures. We show beside the AB$_2$ circuits, figures of equivalent systems which have exactly the same conductance profiles. Parameters: $t_L = t_R = 0.3 t$.}
    \label{fig:conductanceplots}
\end{figure*}
%%%%%%%%%%%%%%%%%%%%% 
%      end          % 
%%%%%%%%%%%%%%%%%%%%%

In Fig.~\ref{fig:conductanceplots}, we show several profiles of the conductance through the AB$_2$ ring with four unit cells as function of the energy of the incident electron (or chemical potential of the leads) or as function of a potential $V_\text{gate}$ applied to the AB$_2$ ring. These profiles correspond to certain positions of the leads which are shown at the top of the  Fig.~\ref{fig:conductanceplots}a, Fig.~\ref{fig:conductanceplots}b and Fig.~\ref{fig:conductanceplots}c. In these figures we also include diagrams of equivalent systems, that is, systems that exhibit exactly the same conductance profiles as the AB$_2$ ring.

In the case of Fig.~\ref{fig:conductanceplots}a, the leads are connected to sites A, therefore the anti-bonding  BC "sites" can be ignored since they are completely decoupled from the leads.
The remaining "ring" of sites A and bonding BC sites form a tight binding ring. 
%%%%%%%%%%%%%%%%%%%%% 
%      figure       % 
%%%%%%%%%%%%%%%%%%%%% 
\begin{figure}[tp]
    \centering
    \includegraphics[width=.4 \textwidth]{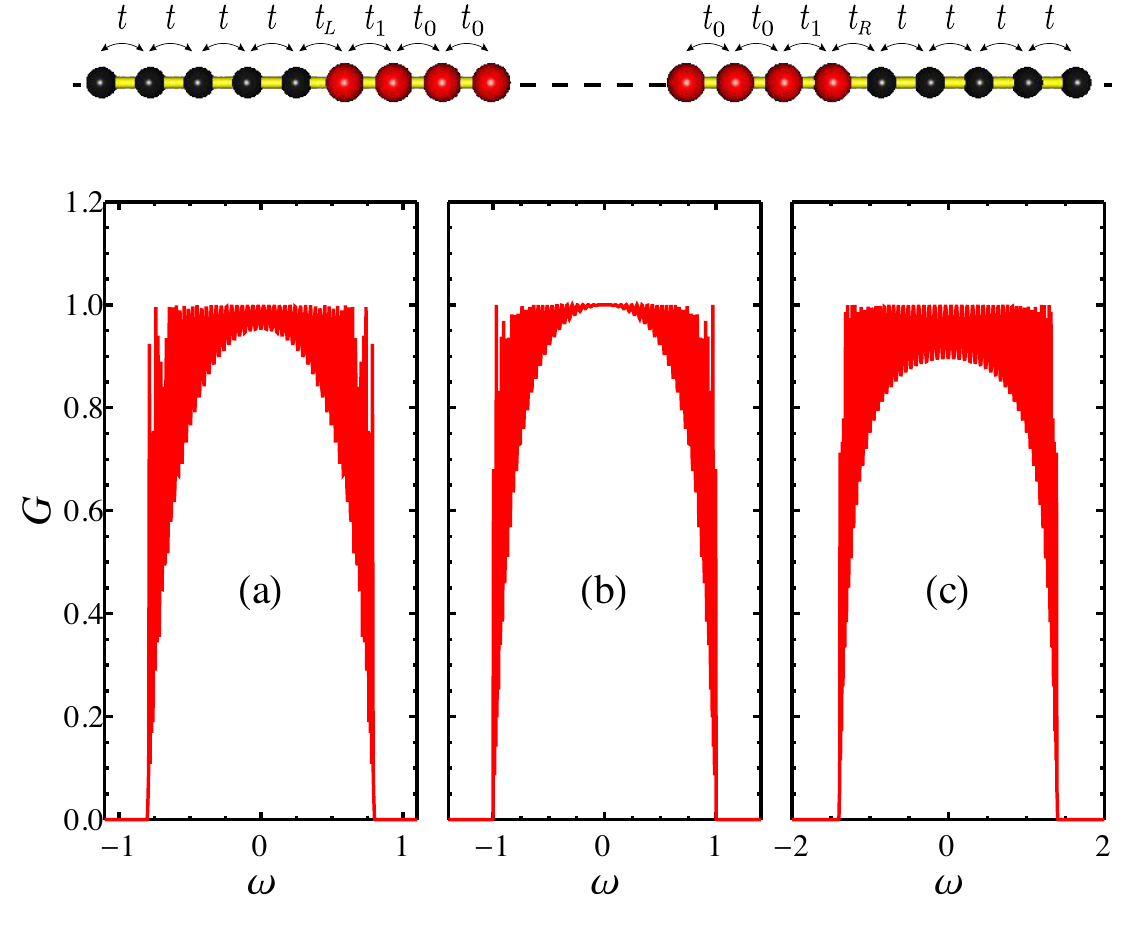} 
    \caption{The conductance through the AB$_2$ ring (or a linear ring) in the absence of flux and with opposite contacts at sites A is the same as for a linear chain (shown in the top figure, where the larger, red sites replace the AB$_2$ ring). In (b), one has the conductance for  the particular case of $2t_1/\sqrt{2}=2t_0=t_L=t_R=1$ in units of $t$ and one sees that  the central cluster (corresponding to the AB$_2$ ring) becomes transparent to an incoming particle for energies around zero, i.e., the bond with hopping constant $t_1$ acts as a monoatomic anti-reflection coating between the regions with hoppings $t$ and  $t_0$. Deviations of the hopping constants from the previous values introduce oscillations in the conductance for energies around zero, as shown in (a) and (c) (where $t_1/\sqrt{2} =t_0=.4$ and $t_1/\sqrt{2} =t_0=.7$, respectively).
}
    \label{fig:monoatomic}
\end{figure}
%%%%%%%%%%%%%%%%%%%%% 
%      end          % 
%%%%%%%%%%%%%%%%%%%%%
Therefore, if the contacts are sites A, the conductance is exactly the same as that of the equivalent tight-binding ring (with hopping constant $\sqrt{2} t$).\cite{kowal_transmission_1990} For small coupling between the leads and the AB$_2$ ring, the conductance has peaks when the chemical potential coincides with any of the system eigenvalues of the AB$_2$ ring, due to resonant tunnelling. These peaks have the Breit-Wigner shape. In  Fig.~\ref{fig:conductanceplots}a, three peaks A, B and C are observed in $G(V_\text{gate})$ in a potential interval corresponding to the bandwidth of the leads (the chemical potential of the leads is equal to zero). The same  peaks should also be observed in the $G(\omega)$-plot of Fig.~\ref{fig:conductanceplots}a where  $V_\text{gate}=0$ and the chemical potential (or equivalently the energy $\omega$ of the incident particle) is varied from the bottom of the leads band to the top. However  peaks A and B are absent  because they correspond to the bottom and  top energies of    the leads bands and the particle velocity is zero for these energies.

If the left contact is a site B or C (let us assume it is a site B) and the right contact is a site A, as in the case of Fig.~\ref{fig:conductanceplots}(b), the conductance profile  is the same as that of a tight-binding ring  but with the $\omega=0$ peak  absent. This absence reflects the fact that the hopping term from the left lead to a B site of the AB$_2$ ring, in the basis of antibonding BC$^-$, bonding BC$^+$ and  A states, becomes a hopping  between the left lead and a  bonding state bonding BC$^+$  and a hopping to a localized state BC$^-$, both with a smaller hopping constant $t_L/\sqrt{2}$. Since this localized state is decoupled from all other states of the ring, it only leads to a reflected wave back into the left lead. For $\omega=0$, this reflected wave interferes destructively with the incident wave and one can say the localized state BC$^-$ acts as a conductance absorber for frequencies close to $\omega=0$ (in close analogy with $\lambda/4$ sound absorbers). The absence of the $\omega=0$ peak can also be explained in the following way. The hopping to the  BC$^-$ "site" is a "dangling bond". If one considers  a linear chain with a dangling site as shown in  Fig.~\ref{fig:figonelink} (with hopping constant $t_a$ to the dangling site), then the  equation for the wavefunction amplitude $\psi_a$  at the dangling site of a particle with energy $\omega=-2t \cos (k)$  is $\omega \psi_a=-t_a \psi_j$. Substituting  $\psi_a$ in the 
the  equation for the wavefunction amplitude at  site $j$, $ \psi_j$, one has 
$\omega \psi_j=-t \psi_{j-1}-t \psi_{j+1}+(t_a^2/\omega) \psi_j$, therefore  the dangling site  effectively modifies the on-site energy of site $j$ to $\epsilon_j(\omega)= (t_a^2/\omega)$. 
When $\omega=0$, the on-site energy becomes infinite and one has zero conductance at $\omega=0$. 
The peaks A and C in Fig.~\ref{fig:conductanceplots}(b) have a reduced amplitude compared to those in Fig.~\ref{fig:conductanceplots}a due to the difference in paths in the upper and lower arms of the ring.

If both the left and right contacts are sites B (or C) as shown in  Fig.~\ref{fig:conductanceplots}(c), an analogous reasoning applies and the system is equivalent to a linear ring connected to leads but with two dangling sites, one at the end of each lead. Again,  localized states act as a filter of the $\omega=0$ peak.

%%%%%%%%%%%%%%%%%%%%% 
%      figure       % 
%%%%%%%%%%%%%%%%%%%%% 
\begin{figure*}[t]
    \centering
	\subfloat[]{\includegraphics[width=.45 \textwidth]{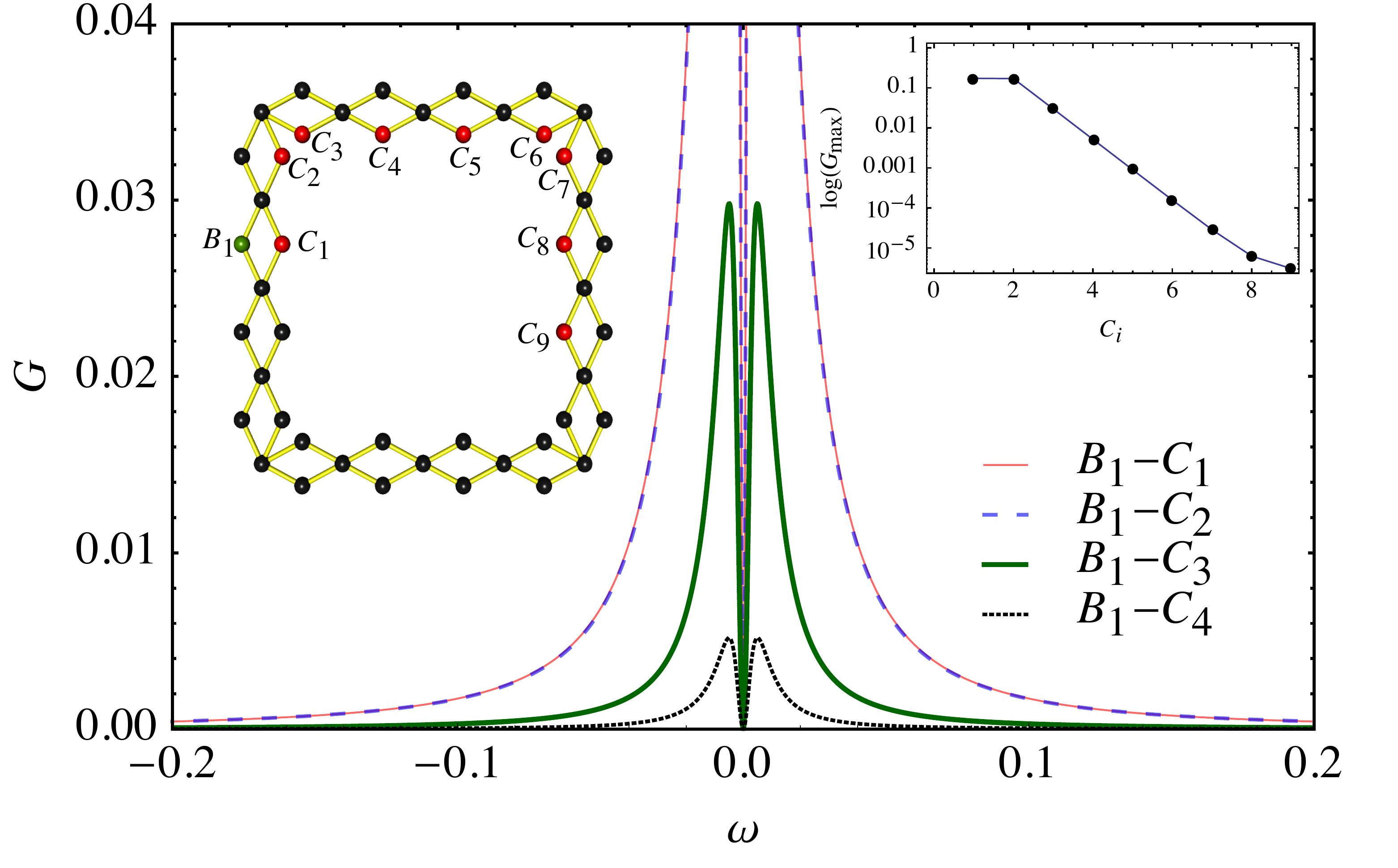}
	\label{fig:fromflattoflat}}
	\subfloat[]{\includegraphics[width=.44 \textwidth]{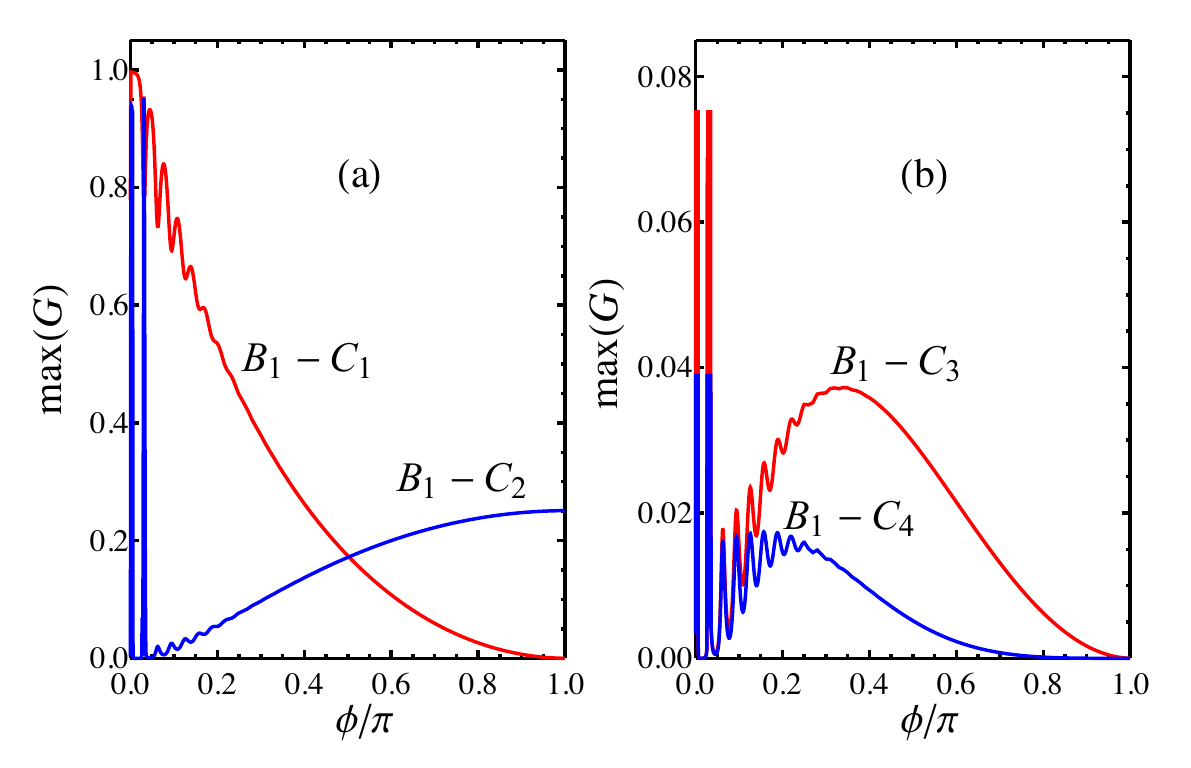}
	\label{fig:figs_Gs}}
    \caption{
    (a) Conductance as a function of the frequency of the incident particle for an AB$_2$ ring of 16 plaquettes,  $\phi = \pi/2$, and several  positions of the leads contacts. For this flux value, only localized states contribute significantly to the $\omega=0$ conductance peak.
    (b) Maximum of the conductance as a function of the flux for 16 plaquettes. For leads contactsat B$_1$-C$_2$ and $\phi =\pi$, the peak value of the conductance is $0.25$, since in this case the AB$_2$ is mapped onto the cluster of Fig.~\ref{fig:flatflat} with equal values of the hopping constants. For leads at B$_1$-C$_1$ it is zero for $\phi =\pi$, since the left and right leads couple to orthogonal states which do not overlap. The oscillations at low flux reflect dependence of the inner magnetic flux $\phi_i$ and disappear as the gap between the itinerant bands and the localized states grows with increasing flux.}
    \label{fig:condloc}
\end{figure*}
%%%%%%%%%%%%%%%%%%%%% 
%      end          % 
%%%%%%%%%%%%%%%%%%%%%

%%%%%%%%%%%%%%%%%%%%% 
%      figure       % 
%%%%%%%%%%%%%%%%%%%%% 
\begin{figure}[ht]
    \centering
    \includegraphics[width=8cm]{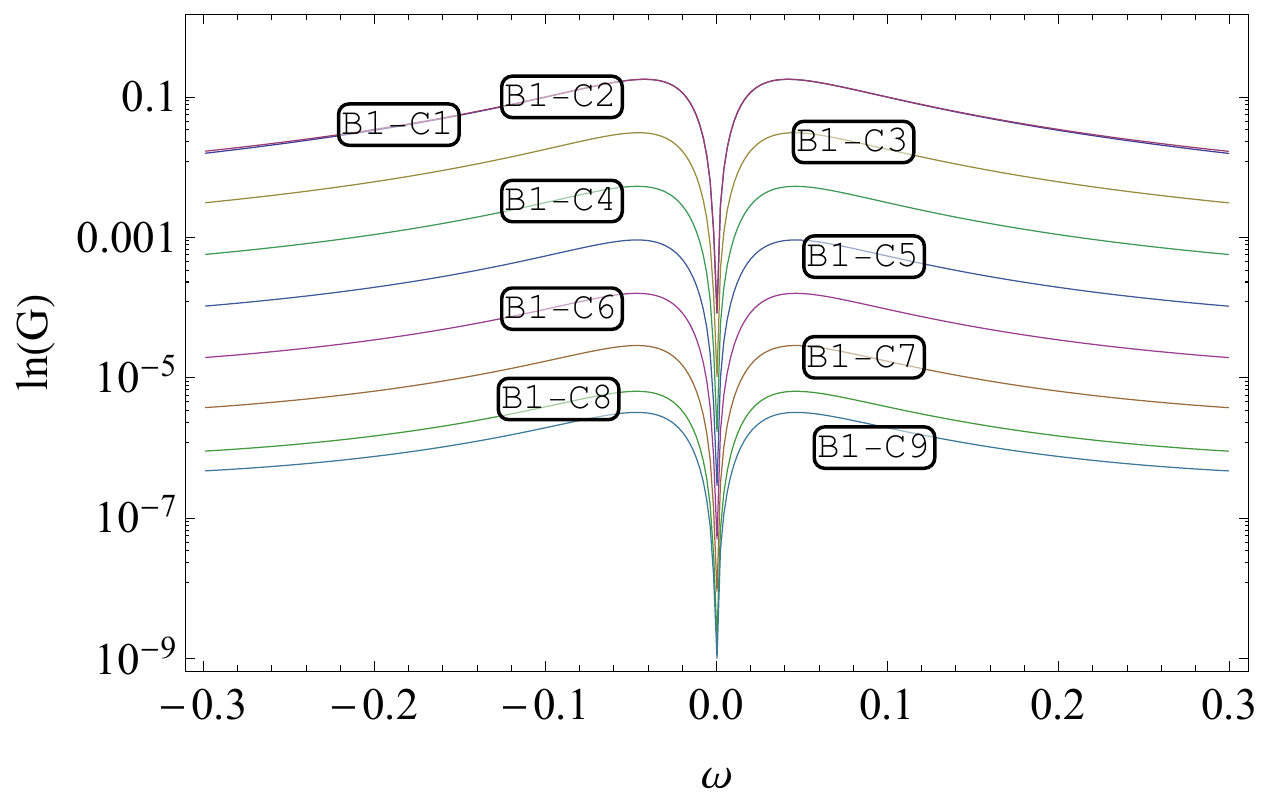}
    \caption{Logarithm of the conductance for 16 cells and $\phi = \pi/2$ around $\omega = 0$, for all the inequivalent lead positions such that one of the leads is on a B site and the other on a C site.}
    \label{fig:logarithmplot}
\end{figure}
%%%%%%%%%%%%%%%%%%%%% 
%      end          % 
%%%%%%%%%%%%%%%%%%%%%

In the case of  Fig.~\ref{fig:conductanceplots}a,
the remaining "ring" of sites A and bonding BC sites can be mapped onto a linear chain since  the leads are coupled to opposite A sites. 
In this case, the leads define an axis of symmetry of the diamond ring and the anti-bonding combinations of an A (or bonding BC) site with the one obtained by reflection in this axis of symmetry are decoupled from the contact sites, or equivalently, the tight-binding hoppings from the  contact sites generate a bonding combination of the  nearest-neighbor bonding BC "sites" and this bonding combination couples only to the bonding combination of A sites. So, for the purpose of calculating the conductance across the AB$_2$ ring,  it is enough to  consider the linear sequence of these bonding states (see the top diagram in  Fig.~\ref{fig:monoatomic}  where the cluster of larger, red sites replaces the AB$_2$ ring). In  Fig.~\ref{fig:monoatomic}b, one has the conductance for  the particular case of $2t_1/\sqrt{2}=2t_0=t_L=t_R=1$ in units of $t$, that is, we have three regions with different hopping constants $t$, $t_0$ and $t$, separated by single hoppings of constant $t_1$. The central cluster (and therefore also the AB$_2$ ring) becomes transparent to the incoming particle for energies around zero. Deviations of the hopping constants from the previous values introduce oscillations in the conductance for energies around zero, as shown in Fig.~\ref{fig:monoatomic}a and Fig.~\ref{fig:monoatomic}c, where the dome of minima of the conductance  oscillations is below one.

This result can be explained with an analogy with a quarter wavelength anti-reflection coating \cite{hecht_optics_2002}, that is, the bond with hopping constant $t_1$ acts as a monoatomic anti-reflection coating between the regions with hopping constant $t$ and  $t_0$. A  anti-reflection coating generates an additional 
reflected wave which is out of phase with the first reflected wave and therefore partially cancels the reflection. If the refraction index of the coating is the geometric mean of the refraction indices of the materials to the left and right of the coating, $n_c=\sqrt{n_\text{left}n_\text{right}}$, the transmittance becomes one when the wavelength of incident wave, $\lambda$, is such that the thickness of the coating is an 
odd multiple of the $\lambda/4$.
This can be translated into our problem in the following way.
The relation $n_c=\sqrt{n_\text{left}n_\text{right}}$ can be written as 
$n_c/n_\text{right}=\sqrt{n_\text{left}/n_\text{right}}$ which is equivalent to a relation between velocities $v_\text{right}/v_c=\sqrt{v_\text{right}/v_\text{left}}$.
The ratio between velocities in our system for energies close to zero is approximately the ratio of hopping constants and the previous relation becomes
$t_1/t_0=\sqrt{t/t_0}$. So perfect transmission occurs when 
$2t_1/\sqrt{2}=2t_0=t$ (as in the case of   Fig.~\ref{fig:monoatomic}b) and when $\lambda/4$ is equal to one interatomic distance (which we have assumed to be one), that is, for $k=\pi/2$ or equivalently, energy $\omega=-2t \cos(k)$ equal to zero.

%%%%%%%%%%%%%%%%%%%%% 
%      figure       % 
%%%%%%%%%%%%%%%%%%%%% 
\begin{figure*}[t]
    \centering
	\subfloat[]{\includegraphics[width=.3 \textwidth]{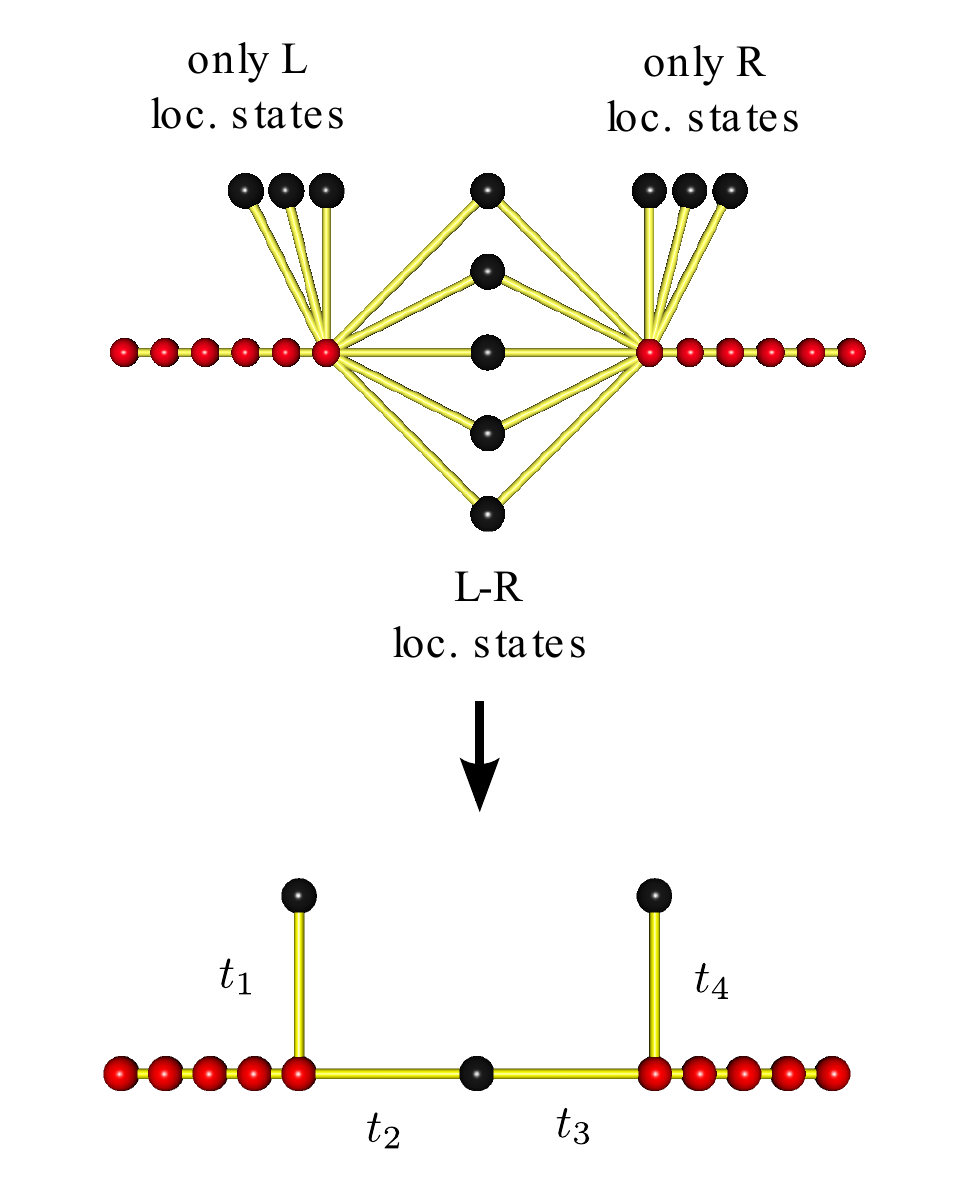}
	\label{fig:flatflat}}
	\quad\quad\quad
	\subfloat[]{\includegraphics[width=.5 \textwidth]{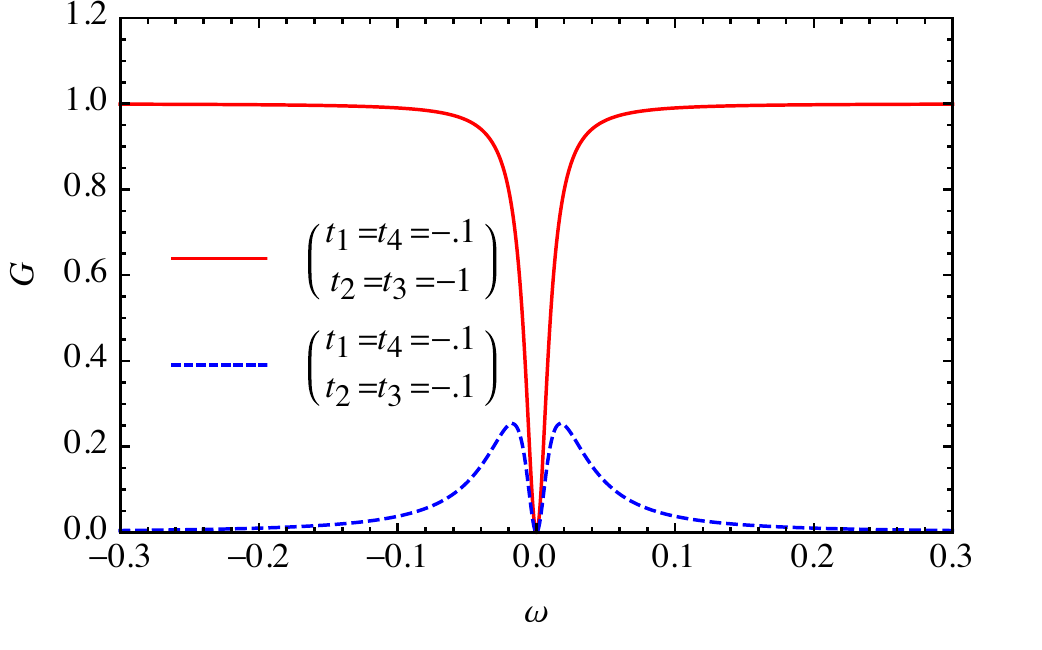}
	\label{fig:figs_noflux_equiv}}
    \caption{(a) The coupling of the leads to the localized states of the AB$_2$ ring in the presence of flux is described by the top diagram where one has dangling sites connected to the left (right) lead representing the localized states which have finite wavefunction amplitude at site L (R), but zero amplitude at site R (L). The system on top  is equivalent to the bottom system  consisting of only one L, R and LR localized states. (b) Whenever all hopping constants in the bottom system are equal, the maximum of the conductance is 0.25 regardless of their value. Due to the dangling sites, the conductance always goes to zero at zero frequency.}
    \label{fig:figs_noflux_equiv2}
\end{figure*}
%%%%%%%%%%%%%%%%%%%%% 
%      end          % 
%%%%%%%%%%%%%%%%%%%%%
When magnetic flux is present, a gap opens between the flat band and the itinerant bands of the AB$_2$ ring.
The conductance peaks corresponding to energies of itinerant states follow the behavior of  the conductance peaks of a linear ring and we do not address them here (see \cite{kowal_transmission_1990}).
The behavior of the conductance  for energies close to $\omega=0$ (which is determined only by the localized states of the AB$_2$ ring when flux is finite and  the coupling between the leads and the cluster is small) is rather unusual.
 A  zero frequency dipped conductance peak is observed despite the fact that the energies of the itinerant states of the ring are far from this zero frequency peak.  This is shown in  Fig.~\ref{fig:fromflattoflat} for the case of an AB$_2$ ring with 16 unit cells and for $\phi=\pi/2$. This dipped peak only occurs if the contact sites are sites B or C (otherwise the peak is absent) and 
shows distinct behavior as function of the positions of the contact sites and as function of the magnetic flux.
As shown in the inset of  Fig.~\ref{fig:fromflattoflat}, the peak maximum decays quasi-exponentially as a function of the contact sites distance, except for the first two distances, where the conductance peak maximum remains the same. Such behavior is also visible in Fig.~\ref{fig:logarithmplot} where the logarithm of the conductance for small frequencies of an incident particle is plotted for several inequivalent positions of the leads.
The dependence with flux of the maximum of the conductance peak shows rather  peculiar behavior depending on the position of the contacts.
In  Fig.~\ref{fig:figs_Gs}, we show the maximum of the $\omega=0$ conductance peak as a function of the flux threading each plaquette, $\phi$, for several choices of contacts positions.
When the contacts are the sites B$_1$ and C$_1$ (see the labeling of the sites in the inset of  Fig.~\ref{fig:fromflattoflat}), the maximum starts at one, oscillates for small $\phi$ and goes smoothly to zero as $\phi$ approaches $\pi$.
The oscillations near $\phi=0$ reflect the contribution to  the conductance of the itinerant states which oscillates as a consequence of the Aharonov-Bohm effect due to the varying flux threading the inner region of the AB$_2$ ring  (an uniform field was applied to the AB$_2$ cluster). These oscillations disappear as $\phi$ grows due to the larger gap between the itinerant bands and the $\omega=0$ energy. Note that  localized states 
do not "feel" this inner flux, that is,  their energy is independent of this field and therefore do not contribute to an Aharonov-Bohm effect.
When the contacts are the sites B$_1$ and C$_2$, contrasting behavior occurs and  the maximum approaches zero when $\phi$ is small and tends to 0.25 when $\phi$ goes to $\pi$.
For larger distance between contacts, the graphs of the maximum of the $\omega=0$ conductance peak exhibit   dome-like profiles (see Fig.~\ref{fig:figs_Gs}(b)), with the peak maximum growing  from near zero when $\phi$ is small, reaching a maximum value and decreasing to zero when $\phi$ approaches $\pi$.

These results can explained recalling our previous discussion of the extension of localized states when the flux is finite. Since this extension ranges from two unit cells to the full ring, and ignoring the itinerant states of the ring which are energetically far from the $\omega=0$ energy region, the conductance is only finite if one has localized states that extend from the left contact to the right contact in the AB$_2$ ring. More precisely, we can divide the localized states in the following way:
states I that extend from the left contact to the right contact, that is, that have finite wavefunction amplitudes at the sites L and R of the AB$_2$ ring; states II that have finite wavefunction amplitudes at the site L but not at the site R of the AB$_2$ ring; states III that have finite wavefunction amplitudes at the site R but not at the site L of the AB$_2$ ring; states IV that have zero wavefunction amplitudes at both the sites L and R of the AB$_2$ ring. Note that the choice of basis for the subspace of localized states influences the number of states in each of these groups, but the explanation for the  conductance results remains the same.
The larger the extension of the localized states, the smaller the wavefunction amplitude at the contact sites, and consequently the smaller the effective hopping between the extremities of the leads and the localized state.
So, our system is equivalent to that displayed in the top diagram of Fig.~\ref{fig:flatflat}.
%, but again depending on the choice of the basis, for example,   states that have finite amplitude only at site R may not exist.
The hopping constants  between the leads (smaller, red spheres) and the localized states (larger, black spheres) are in general  different but all these hopping can be simplified and the system can be reduced to that shown in the bottom diagram of Fig.~\ref{fig:flatflat}. In fact,  several dangling sites (states II) contribute to the on-site energy of the site at the end of the left lead and $t_1$ is the hopping constant that generates an on-site energy equal to the sum of the on-site energies generated by the dangling sites at the left lead. The same goes for $t_4$. The effect of the localized states of the type I can also be reproduced with a single site but with different hopping constants to the left lead and to the right lead. In Fig.~\ref{fig:figs_noflux_equiv}, we show the conductance through this simplified system. If $t_2=t_4=t$, without the dangling sites, we would have perfect transmittance  for any energy of the incident electron. The effect of the dangling sites is the creation of the dip at $\omega=0$ as one can see in Fig.~\ref{fig:figs_noflux_equiv} (red solid curve). If $t_2$ and $t_3$ are rather smaller than $t$ and no dangling site is present ($t_1=t_4=0$), a peak appears at $\omega=0$ of width proportional to $t_2$ (assuming $t_2=t_3$).
The effect of the dangling sites in this case is again the introduction of the dip at the center of this peak. If the width of the dip becomes larger than that of the peak, the  dipped peak maximum becomes small. 

One can now explain the behavior displayed in Fig.~\ref{fig:figs_Gs}.
One should recall that the non-orthogonal localized states are of the form 

$
	(\ket{B_j}
	- e^{i\frac{\phi}{2}} \ket{C_j})
	+ (e^{i\frac{\phi}{2}} \ket{B_{j+1}}
		- \ket{C_{j+1}})
$.  
The overlap between consecutive localized states is equal to $\cos(\phi/2)/2$, so it is zero whenever $\phi =\pi$, and $1/2$ when $\phi =0$ (the latter value implies that shorter and orthogonal localized states can be found of the form 
BC$_j^-= 1/\sqrt{2 } ( \ket{B_{j}}- \ket{C_j} )$).
We consider only the mean evolution of the conductance, that is, the  dependence of the conductance remaining if the oscillations due to the Aharonov-Bohm effect are removed. This behavior consists of the following:
if the contacts are the sites B$_1$ and C$_1$, the maximum of the conductance is one 
for zero flux and with increasing magnetic flux, the conductance decreases and becomes zero for flux equal to $\pi$.
Note that for $\phi=\pi$, the localized states, $ \ket{\psi_j} =	1/2 \ket{(B_j}
	- i \ket{C_j})
	+ (i \ket{B_{j+1}}
		- \ket{C_{j+1}} $, 
are orthogonal 
and  both leads couple to only two of these states, 
$ \ket{\psi_1} $
and 
$ \ket{\psi_{N_c}} $. That is, we have only two states of type I and all other localized states are of type IV.
The transport through the cluster is given by the transfer terms to these localized states of the form $\bra{\psi_1} H \ket{0}$ which collected (omiting the hopping terms in the leads) give rise to 
$
(-t_L/2) \ket{0} [ \bra{\psi_1} 
+
 i  \bra{\psi_{N_c}}]  
+
(-t_R/2) \ket{N+1}  [  -i  \bra{\psi_1} 
-
 \bra{\psi_{N_c}}] 
+
\text{H.c.}
$
But 
$[ \bra{\psi_1} 
+
 i  \bra{\psi_{N_c}}]$ 
and 
 $[  -i  \bra{\psi_1} 
-
 \bra{\psi_{N_c}}]$ 
 are orthogonal bras, therefore the left and right leads are effectively decoupled and the transmittance is zero.
A similar reasoning can be followed when $\phi$ approaches zero. In this case the leads couple to only one localized state, 
BC$_1^-= 1/\sqrt{2 }( \ket{B_{1}} - \ket{C_1})$,
that is, we have one state of type I and no dangling sites, so the transmittance approaches one.

If the contacts are the sites B$_1$ and C$_2$, the maximum of the conductance is zero 
for zero flux and, with increasing magnetic flux, the conductance increases and becomes $1/4$ for flux equal to $\pi$. The fact that the conductance maximum approaches zero as $\phi $ goes to zero is common to all other contact possibilities with exception of the previous one and reflects a similar  argument, that is, the left lead couples to only one localized state, 
BC$_1^-= 1/\sqrt{2 }( \ket{B_{1}} - \ket{C_1})$
and the right lead couples only to one other localized state which is orthogonal to the former, and consequently the transmittance is zero.
The fact the conductance goes to $1/4$ when the flux goes to $\pi$ can also be justified as before,  collecting the transfer integrals
and one has
$
(-t_L/2) \ket{0} [ \bra{\psi_1} 
+
 i  \bra{\psi_{N_c}}]  
+
(-t_R/2) \ket{N+1}  [  -  \bra{\psi_1} 
-i
 \bra{\psi_2}] 
+
\text{H.c.}
$,
and this corresponds to the bottom diagram displayed in Fig.~\ref{fig:figs_noflux_equiv} with $t_1=it_L/2$, $t_4=-it_L/2$, $t_2=t_L/2$, and $t_3=t_R/2$. Since we considered $t_L=t_R$, all these hopping constants are equal in absolute value and therefore  the conductance   is equal to 1/4 in agreement with what is shown in  Fig.~\ref{fig:figs_noflux_equiv}. Note that  the phase terms are irrelevant at the dangling sites.

If one of the contacts is the site B$_1$ and the other is a C$_j$ site with $j \neq N_c,1,2$, the maximum of the conductance goes to zero 
as the flux goes to zero and with increasing magnetic flux, the conductance increases, reaches a maximum (this maximum becomes smaller as the distance between contacts increases)  and goes again to  zero when the flux approaches  $\pi$, reflecting the fact that the orthogonal localized states
are all two unit cells long.

\section{Conclusions}

We have shown that localized states in itinerant geometrically frustrated electronic systems generate rather striking behavior in the two terminal electronic conductance. In the absence of magnetic flux, the localized states act as a filter of the zero frequency conductance peak (we suggested an analogy with $\lambda/4$ sound absorbers), if there is a finite  hopping probability between the leads contact sites and  the localized states. In contrast, when magnetic flux is present, some localized states contribute to the appearance of a zero frequency conductance peak while other localized states act as a conductance absorber, and as a consequence, the conductance exhibits a zero frequency peak with a dip.

We have shown that such different roles of the localized states are due to the fact that the presence of magnetic flux implies that any orthogonal basis of the subspace of localized states 
is composed of localized states with  variable extensions (ranging from two unit cells to the complete ring, in the case of the AB$_2$ ring studied in this paper).
Such peculiar dipped peak fixed at the localized states energy, even when magnetic flux is varied, is a distinct fingerprint of the existence of localized states in itinerant geometrically frustrated electronic systems. Furthermore, depending on the  distance between  contact sites, different profiles for the  maximum of the dipped conductance peak as function of the magnetic flux have been obtained, and this implies that the two terminal conductance  can be used as a probe of the localized states spatial dependence.

%%%%%%%%%%%%%%%%%%%%%%%%%%%%%%%%%%%%%%%%%%%%%%%%%%%%%%%%%%%%%%%%%%%%%%%%%%
%                        Acknowledgements                                 %
%%%%%%%%%%%%%%%%%%%%%%%%%%%%%%%%%%%%%%%%%%%%%%%%%%%%%%%%%%%%%%%%%%%%%%%%%%

\subsection*{Acknowledgements}
A. A. Lopes acknowledges the financial support of the Portuguese Science
and Technology Foundation (FCT), cofinanced by FSE/POPH, under grant SFRH/BD/68867/2010 and of the Excellence Initiative of the German Federal and State Governments (grant ZUK 43).
R. G. Dias  acknowledges the financial support from the Portuguese Science
and Technology Foundation (FCT) through the program PEst-C/CTM/LA0025/2013.

\bibliography{prb}

\end{document}